\documentclass[smallextended]{svjour3my}
\smartqed
\usepackage{graphicx}

\usepackage{hyperref}
\journalname{Gen Relativ Gravit}
\begin{document}


\title{Number of revolutions of a particle around a black hole:
Is it infinite or finite?}

\titlerunning{Number of revolutions of a particle around a black hole...}

\author{Yuri V. Pavlov$^{1,2}$        \and
        Oleg B. Zaslavskii$^{2,3}$}

\authorrunning{Yu.V. Pavlov \and O.B. Zaslavskii}

\institute{Yu. V. Pavlov \at
              \email{yuri.pavlov@mail.ru}
           \and
           O. B. Zaslavskii \at
              \email{zaslav@ukr.net}
           \and
${}^{1}$\,Institute of Problems in Mechanical Engineering,
Russian Academy of Sciences,\\
61~Bol'shoy pr., St. Petersburg 199178, Russia
           \\ \\
${}^{2}$\,N.I.\,Lobachevsky Institute of Mathematics and Mechanics,
    Kazan Federal University, 18 Kremlyovskaya St., Kazan 420008, Russia;
           \\ \\
${}^{3}$\,Department of Physics and Technology, Kharkov V.N.Karazin
National University,~4 Svoboda Square, Kharkov 61022, Ukraine}

\date{Received: 12 July 2017 / Accepted: 14 December 2017}

\maketitle

\begin{abstract}
We consider a particle falling into a rotating black hole. Such a particle
makes an infinite number of revolutions $n$ from the viewpoint of a remote
observer who uses the Boyer-Lindquist type of coordinates. We examine the
behavior of $n$ when it is measured with respect to a local reference frame
that also rotates due to dragging effect of spacetime. The crucial point
consists here in the observation that for a nonextremal black hole, the
leading contributions to $n$ from a particle itself and the reference frame
have the same form being in fact universal, so that divergences mutually
cancel. As a result, the relative number of revolutions turns out to be
finite. For the extremal black hole this is not so, $n$ can be infinite.
Different choices of the local reference frame are considered, the results
turn out to be the same qualitatively. For illustration, we discuss two
explicit examples --- rotation in the flat spacetime and in the Kerr metric.

\end{abstract}

\section{Introduction}

\label{intro}

It is well known that for the Kerr metric~\cite{kerr} in the Boyer-Lindquist
coordinates~\cite{bl}, a remote observer finds that the number of
revolutions around the area bounded by the event horizon tends to infinity
when the horizon is approached~\cite{ch}. In doing so, the angular velocity
of a particle approaches that of a black hole, while the time measured by a
distant observer tends to infinity. However, the proper time needed to reach
a black hole is typically finite and, generally speaking, even quite modest.
Then, a natural question arises, how measurements performed by both
observers are related to each other. And, is it possible to perform an
infinitely large number of revolution (hence, cover an infinite proper
distance) during a finite interval of the proper time? In some exceptional
situations (so-called critical particles with fine-tuned parameters in the
background of the extremal black hole) a corresponding proper time is
infinite and this also requires special attention.

According to basics of general relativity, for the description of any region
of spacetime, any mathematically correct coordinate systems can be
exploited. Therefore, the answer to the question about finite or infinite
number of revolutions in the fall to a black hole may depend on the way
observations are performed. For instance, there are two typical observers -
a remote observer residing in a region of weak field far from a black hole
and a cosmonaut falling into a black hole. Moreover, even for a remote
observer itself, the effect considered below depend on a frame strongly.

Before the analysis of the case of a rotating black hole we will consider a
more simple one of rotation in the flat spacetime and discuss, whether the
paradoxical situations under discussion can arise there.

\section{Rotation in flat spacetime\label{flat}}

Let a massive particle move in the Minkowski spacetime with the metric
\begin{equation}
ds^{2}=-c^{2}dt^{2}+dx^{2}+dy^{2}+dz^{2} ,
\label{mink}
\end{equation}
where $c$ is the light velocity.

\subsection{Is infinite distance consistent with a finite proper time for
linear motion?}

First of all, let us consider motion of a particle (not necessary a free
one) along a straight line. Whether a particle can cover an infinite proper
distance for a finite time? It is obvious that a coordinate time $\Delta t$
necessary for such a travel is infinite. What can be said about the interval
of the proper time $(\tau_{1},\tau_{2})$?

We assume that a particle moves along the $x-$axis. Then, for motion between
points $(t_{1},x_{1})$ and $(t_{2},x_{2})$ we have
\begin{equation}
x_{2}-x_{1}=\int\limits_{t_{1}}^{t_{2}}v\,dt=\int\limits_{\tau
_{1}}^{\tau_{2}} \frac{v\,d\tau }{\sqrt{1-\frac{v^{2}}{c^{2}}}},
\label{pl2}
\end{equation}
where $v$ is the velocity. Taking into account that
\begin{equation}
\sqrt{1-\frac{v^{2}}{c^{2}}}=\frac{mc^{2}}{E},
\label{pl3}
\end{equation}
where $m$ is the rest mass, $E$ is the energy, we obtain
\begin{equation}
x_{2}-x_{1}=\frac{1}{mc}\int\limits_{\tau_{1}}^{\tau_{2}}
\sqrt{ E^{\mathstrut 2}-m^{2}c^{4}} \, d\tau .
\label{pl4}
\end{equation}

Thus if we want the proper distance to diverge with the finite interval of
the proper time, the energy~$E$ should grow unbounded within this interval,
so that the velocity should approach that of light. Hence, a traveller who
wants to see an infinite distance for a finite time, should have an infinite
fuel supply that is, obviously, impossible.

A similar problem arises because of the behavior of acceleration. For motion
with a constant proper acceleration $a$ distances covered for any finite
interval of the proper time $\tau_{1}<\tau <\tau_{2}$ are also finite (see
e.g. the problem after Sec. 7 of Chapter 1 in \cite{LL}). If an infinite
distance is covered during a finite $\Delta \tau =\tau_{2}-\tau_{1}$, $a$
should grow unbounded when $\tau \rightarrow \tau_{2}$. Otherwise, if
$ a<a_{0}<\infty $, motion with any constant $a>a_{0}$ would increase the
proper distance but, nonetheless, it would remain finite according to the
aforementioned property. This causes contradiction, so an infinite proper
distance is impossible for the bounded $a$. Additionally, any material body
cannot withstand unbounded proper acceleration.

\subsection{Relative number of revolutions for a fixed radial distance}

    Does this forbid the possibility to perform an infinite number of
revolutions around a black hole for a finite interval of the proper time?
    In the black hole case the above arguments do not work.
    Although the proper acceleration of a free falling particle is equal to zero,
its velocity with respect to a stationary frame tends to that of light when a particle
approaches a black hole horizon~\cite{LL}.
    Also, in particle collisions near the horizon the energy in the
centre of mass may grow unbounded~\cite{bsw}, \cite{prd}, \cite{gp11}, hence
their relative velocity tends to speed of light~\cite{k}.
    It was shown in~\cite{gp16} that there is a lot similarities between motion
of particles in the rotating coordinate frame in the flat spacetime and free fall of
particles in the Kerr metric in the Boyer-Lindquist coordinates.
    In the rotating coordinates in Minkowski space-time and in Kerr metric there are
regions where particles cannot be at rest: the ergosphere in Kerr metric and
the region out of the surface $r=c/\Omega $ ($\Omega \geq 0$ is the angular
velocity of rotation) in rotating system.
    States of particles with negative and zero energies are possible in these regions.
    The angular velocity of particles in these regions has always the same direction
and is limited between the maximal and minimal values.
    Inside the ergosphere and in the region $r>c/\Omega $ in rotating system the value
of the particle angular velocity defines the sign of the particle energy.
    There is also an analogy between the event horizon of the Kerr black hole and
radial space infinity of the rotating coordinate system.
    In both cases it takes an infinite coordinate time to reach them and the angular
velocity is going to a definite limit: to the angular velocity of the rotation of
the black hole or to the angular velocity of the coordinate system.
    For these reasons, it makes sense to use the flat metric as a simplified example
to gain insight into what happens in a more complicated case.
    Here, we will consider the question about the number of revolutions.

It is convenient to rewrite the metric~(\ref{mink}) in the cylindrical
coordinates
\begin{equation}
ds^{2}=-c^{2}dt^{2}+dr^{\prime 2}+r^{2}d\phi ^{\prime 2}+dz^{\prime 2}.
\label{mink2}
\end{equation}
Passing to the rotating cylindrical coordinates
\begin{equation}
r^{\prime }=r,\ \ \ \ z^{\prime }=z,\ \ \ \ \phi ^{\prime }=\phi -\Omega t,
\label{v2}
\end{equation}
one obtains
\begin{equation}
ds^{2}=-(c^{2}-\Omega ^{2}r^{2})\,dt^{2}-2\Omega r^{2}d\phi
\,dt+dr^{2}+r^{2}d\phi ^{2}+dz^{2}.  \label{v3}
\end{equation}

We make use the following notations for the integrals of motion:
$ \varepsilon = E/(mc^{2})\geq 1$ is the specific energy in the nonrotating
coordinate frame, $p_{z}$ is the projection of the momentum on the $z$-axis,
$L_{z}$ being the projection of the angular momentum on this axis.
    Then, equations of motion for a particle of a given mass $m$ in the rotating
coordinate frame can be written in the form
\begin{equation}
\frac{dt}{d\tau }=\varepsilon ,\ \ \ \ \frac{dz}{d\tau }=\frac{p_{z}}{m},\ \
\ \ \frac{d\phi }{d\tau }=\frac{L_{z}}{mr^{2}}+\Omega \varepsilon ,
\label{pl5}
\end{equation}
\begin{equation}
\frac{dr}{d\tau }=\pm c\,\sqrt{\varepsilon^{2}-1-\left( \frac{p_{z}}{mc}
\right) ^{2}-\left( \frac{L_{z}}{mcr}\right) ^{2}}.
\label{pl6}
\end{equation}
From eqs.~(\ref{pl5}), (\ref{pl6}), one obtains the equation of a trajectory
in terms of the coordinates~$r$, $\phi$\,:
\begin{equation}
c\frac{d\phi }{dr}=\pm \frac{\frac{L_{z}}{mr^{2}}+\Omega \varepsilon }{\sqrt{
\varepsilon ^{2}-1-\left( \frac{p_{z}}{mc}\right) ^{2}-\left( \frac{L_{z}}{
mcr}\right)^{2}}}.
\label{1}
\end{equation}
Let a particle move with increasing of the radial coordinate from the point
with the coordinate $r_{0}$ to the point with the coordinate $r>r_{0}$. For
definiteness, we choose sign ``+'' in eq.~(\ref{1}), then
\begin{eqnarray}
&&\phi (r)-\phi (r_{0}) =\left( \arcsin \frac{L_{z}}{mcr\sqrt{\varepsilon
^{2}-1-\left( p_{z}/mc\right)^{2}}}\right. - \\
&&-\left. \left. \frac{\varepsilon \Omega r/c}{\varepsilon^{2}-1-\left(
p_{z}/mc\right)^{2}}\sqrt{\varepsilon^{2}-1-\left( \frac{p_{z}}{mc}\right)
^{2}-\left( \frac{L_{z}}{mcr}\right)^{2}}\right) \right\vert_{r}^{r_{0}}.
\label{2}
\end{eqnarray}
Letting in~(\ref{2}) the value of $r$ grow unbounded, we obtain that also
$\phi (r)\rightarrow \infty $.

Thus in the rotating coordinate system the number of revolutions of a
particle moving to infinity tends to infinity as well. This is a natural
result since for an infinite time the rotating coordinate system performs an
infinite number of revolutions, whereas in the static frame a free particle
moves along a straight line.

However, some conclusions that follow from~(\ref{2}) are not so obvious. Let
us consider two particles 1 and 2 having different parameters and the
corresponding numbers of revolutions $n=(\phi (r)-\phi (r_{0}))/(2\pi )$.
When $r\rightarrow \infty $,
\begin{equation}
n_{1}-n_{2}\approx \frac{\Omega }{c}r \left(\frac{1}{\beta_{2}}-
\frac{1}{ \beta_{1}} \right),
\end{equation}
where
\begin{equation}
\beta =\sqrt{1-\frac{1}{\varepsilon^{2}}-\left( \frac{p_{z}}{\varepsilon mc}
\right)^{2}}=\frac{\sqrt{v_{x}^{2}+v_{y}^{2}}}{c}.  \label{b}
\end{equation}
Here, $v_{x}$ and $v_{y}$ are the corresponding components of velocity in
the static coordinate system. If $\beta_{1}\neq \beta_{2}$,
$ n_{1}-n_{2}\rightarrow \infty $.

Returning to the issue of particles falling into rotating black holes, it is
worth noting that in this case there is no analogue of a static frame. In
principle, one can pass to coordinates comoving with a falling observer but
this is impossible in practice because of the lack of exact analytic
solutions. Instead, one can choose a frame which is not comoving as a whole
but in which the time coordinate represents a local proper time for
free-falling observers on a set of simple
trajectories~\cite{dol}, \cite{nat}, \cite{near}.
The analysis of revolutions in which particular frames is an
interesting separate issue but here we use a more general approach.

\subsection{Relative number of revolutions for a fixed time interval}

In the above treatment, we discussed the properties of $n$ for given
initial~$r_{0}$ and final positions $r$ of both particles.
Meanwhile, it makes sense
to consider a somewhat different problem. Let both particles be at the
initial position $r_{0}$ at time $t=0.$ What can be said about their
relative number of revolutions for a fixed time $t$ when $t\rightarrow
\infty $\,?

It follows from eqs.~(\ref{pl5}), (\ref{pl6}) that
\begin{equation}
\frac{d\phi }{dt}=\frac{L_{z}}{\varepsilon mr^{2}}+\Omega ,
\end{equation}
\begin{equation}
\frac{dr}{dt}=\pm c\,\sqrt{\beta^{2} - \left( \frac{L_{z}}{\varepsilon mcr }
\right)^{2}}.
\end{equation}
Then,
\begin{equation}
\phi =\pm \int \frac{L_z \, dr}{\varepsilon m c r^{2}\sqrt{\beta^{2}-\left(
\frac{L_{z}}{ \varepsilon mcr} \right)^{2}}}+ \Omega t ,  \label{fix}
\end{equation}
\begin{equation}
\phi (t) - \phi (t_0) = \left( \mp \left. \arcsin \frac{L_z}{\varepsilon m c
\beta r(t)} \right) \right|_{t_0}^t + \Omega (t-t_0) .
\label{fixdob1}
\end{equation}
When $t \rightarrow \infty $,
\begin{equation}
r\approx \beta c t,  \label{rb}
\end{equation}
so, $r \rightarrow \infty $ as well. However, the first term
in~(\ref{fix}), (\ref{fixdob1}) remains finite.
The second term diverges but it is the same for each particle.
Therefore, if we compare the motion of two particles, the diverging terms
cancel and $n_{1}-n_{2}$ remains finite ($ n_{1}-n_{2}\le \pi$).

We see that the notion of number revolutions is a quite subtle thing and one
should specify clearly the conditions under which it is calculated.

\section{Basic equations of particle motion in curved background}

Hereafter, we use the system of units, where gravitational constant and the
light velocity are equal to one: $G=c=1$.

Let us consider the axially symmetric stationary metric
\begin{equation}
ds^{2}=-N^{2}dt^{2}+g_{\phi }(d\phi -\omega dt)^{2}+\frac{dr^{2}}{A}
+g_{\theta }d\theta^{2} .
\end{equation}
By assumption, the metric coefficients do not depend on $t$ and $\phi $.
Correspondingly, the energy $E=-mu_{\mu }\xi^{\mu }=-mu_{t}$ and angular
momentum $L=mu_{\mu }\eta^{\mu }=mu_{\phi }$ are conserved.
Here, $\xi^{\mu} $ and $\eta^{\mu }$ are the Killing vectors responsible for
time translation and rotation around the polar axis, respectively,
$u^{\mu}$ the four-velocity.
We assume that the metric coefficients depend on $\theta $
through $\cos^{2}\theta $. Then, it is easy to show that a particle that
starts to move within the plane $\theta =\pi / 2$ will remain in this plane.
When considering generic dirty (surrounded by matter) black holes, we
restrict ourselves to such a motion only. Then, it follows from the geodesic
equations and normalization condition $u_{\mu} u^{\mu }=-1$ that
\begin{equation}
m\dot{t}=\frac{X}{N^{2}} ,  \label{t}
\end{equation}
\begin{equation}
m\dot{\phi}=\frac{L}{g_{\phi }}+\frac{\omega X}{N^{2}} ,  \label{phi}
\end{equation}
\begin{equation}
m\dot{r}=\sigma Z ,  \label{r}
\end{equation}
\begin{equation}
X=E-\omega L ,
\end{equation}
\begin{equation}
Z=\sqrt{X^{2}-N^{2}\left( m^{2}+\frac{L^{2}}{g_{\phi }} \right)} .  \label{Z}
\end{equation}
Here $\sigma =\pm 1$, dot denotes derivative with respect to the proper time
$\tau $. We consider a black hole metric, so that $N=0$ corresponds to the
event horizon $r=r_{+}$.

It follows from~(\ref{t}), (\ref{phi}) and (\ref{r}) that
\begin{equation}
\frac{d\phi }{dt}=\omega +\frac{LN^{2}}{g_{\phi }X} ,  \label{om}
\end{equation}
\begin{equation}
\frac{dr}{dt}=\frac{\sigma ZN^{2}}{X} .  \label{rt}
\end{equation}
It is seen from~(\ref{om}) and (\ref{rt}) that if a particle starts from
$ r=r_{0}$ and moves towards $r_{1}<r_{0}$,
\begin{equation}
\phi =\int_{r_{1}}^{r_{0}} \! \left(\frac{\omega X}{ZN^{2}}+\frac{L}{g_{\phi
}Z} \right) dr,  \label{fr}
\end{equation}
where we assumed that $\phi (r_{0}=r_{1})=0$ and put $\sigma =-1$. Then, the
number of revolution when a particle travels from point 1 to point 0 is
equal to $n=\phi /2\pi $.

As usual, we assume the forward-in-time condition $dt / d\tau >0$, whence
\begin{equation}
X\geq 0 .  \label{x}
\end{equation}
Outside the horizon we have $N>0$. On the horizon itself equality
in~(\ref{x}) is possible.
From~(\ref{r}) we obtain the proper time between points $r$ and $r_{+}$
\begin{equation}
\tau =\int_{r_{+}}^{r}\frac{mdr}{Z} .  \label{pr}
\end{equation}
The quantity $d \phi / dt \equiv \Omega $ is allowed to remain in some
nonzero interval
\begin{equation}
\Omega_{-}\leq \Omega \leq \Omega_{+} ,  \label{int}
\end{equation}
where for equatorial motion
\begin{equation}
\Omega_{\pm }=\omega \pm \frac{N}{\sqrt{g_{\phi }}} .  \label{pm}
\end{equation}
The inequality~(\ref{int}) can be derived from the requirement that the
trajectory for $r=\mathrm{const}$ be timelike or lightlike, so that
\begin{equation}
g_{00}dt^{2}+2g_{0\phi }d\phi dt+g_{\phi }d\phi^{2}\leq 0.  \label{ineq}
\end{equation}
For trajectories with $\dot{r}\neq 0$, the term with $g_{rr}\dot{r}^{2}$
in~(\ref{ineq}) decreases $\Omega_{+}$ and increases $\Omega_{-}$.
Inside the ergoregion, $\Omega_{-}>0$, so a particle should be in orbital
motion there.

\section{Classification of particles}

We call a particle critical if
\begin{equation}
X_{H}=E-\omega_{H}L=0
\end{equation}
and usual if $X_{H}>0$.
Hereafter, subscript ``$H$'' denotes quantities calculated on the horizon.
It is seen from~(\ref{rt}) that $t\rightarrow \infty $ if a particle reaches
the horizon.
Meanwhile, in some cases a particle cannot do it.

Near the horizon,
\begin{equation}
\omega =\omega_{H}-B_{1}(r-r_{+})+B_{2}(r-r_{+})^{2}+ \ldots  \label{omega}
\end{equation}
where $B_{1}>0$ for the Kerr metric,
\begin{equation}
X=X_{H}+B_{1}L(r-r_{+})+O((r-r_{+})^{2}) + \ldots  \label{exp}
\end{equation}
(see~\cite{dirty} for more details).

For the critical particle, condition~(\ref{x}) entails
\begin{equation}
L>0.  \label{lp}
\end{equation}

\section{Nonextremal black hole}

By definition, this means that near the horizon
\begin{equation}
N^{2}\approx 2\kappa (r-r_{+}) ,  \label{nn}
\end{equation}
where $\kappa $ is the so-called surface gravity that is a constant on the
horizon.

It follows from~(\ref{nn}) that the critical particle cannot approach the
horizon of the nonextremal black hole since otherwise one would have
$ X^{2}=O(r-r_{+})^{2}$ and, according to~(\ref{Z}), (\ref{nn}), $Z^{2}<0$.
For a usual one, it is easy to obtain from~(\ref{om}) that this is possible.
Then, it is seen from~(\ref{Z}) that $Z_{H}\approx X_{H}=O(1)$ for a usual
particle. It follows from~(\ref{fr}), $n\rightarrow \infty $
\begin{equation}
n\approx \frac{\omega_{H}}{4\pi \kappa }\left\vert \ln
(r_{1}-r_{+})\right\vert ,  \label{nd}
\end{equation}
when $r_{1}\rightarrow r_{+}$.

\subsection{Relative number of revolutions\label{rel}}

Now, let us consider the numbers of revolutions $n_{1,2}$ for two different
particles~1 and~2. We pose the question, whether the difference $n_{1}-n_{2}$
is finite or infinite.
For each particle, the angle is given by eq.~(\ref{fr}).
Thus the difference
\begin{equation}
n_{1}-n_{2}=\frac{1}{2\pi }\int_{r_{1}}^{r_{0}} \left[ \frac{\omega }{N^{2}}
\left( \frac{ X_{1}}{Z_{1}}-\frac{X_{2}}{Z_{2}} \right) + \frac{1}{g_{\phi }}
\left( \frac{L_{1}}{Z_{1}}- \frac{L_{2}}{Z_{2}} \right) \right] dr .
\label{12}
\end{equation}
The second term is obviously finite here. In the first term, we can write
for two generic usual particles
\begin{equation}
\frac{X_{1}}{Z_{1}}-\frac{X_{2}}{Z_{2}} = \frac{N^{2}}{2} \left( \frac{
m_{1}^{2} + \frac{L_{1}^{2}}{g_{\phi }}}{X_{1}^{2}}-\frac{m_{2}^{2}+ \frac{
L_{2}^{2}}{ g_{\phi }}}{X_{2}^{2}} \right)+O(N^{3}) .
\label{xx}
\end{equation}

As \ a result, divergent terms for each particles mutually cancel and
$ n_{1}-n_{2}$ remains finite in the horizon limit.
One can say that diverging
contribution from rotation of a space itself mutually cancels for both
particles since, according to~(\ref{nd}), it does not depend on their
characteristics.

\subsection{The maximum possible relative number of revolutions}

It follows from~(\ref{om}), (\ref{int}) that for any massive geodesic
particle
\begin{equation}
L_{-}<L<L_{+} , \ \ \ L_{\pm }=\pm \frac{X}{N}\sqrt{g_{\phi }} .
\label{maxmin}
\end{equation}
Then, one can derive the upper bound on $n_{1}-n_{2}$ in~(\ref{12}) by
substitution $L=L_{+}$ for particle 1 and $L=L_{-}$ for particle 2.

As a result,
\begin{equation}
n_{1}-n_{2}<n_{\max }^{(12)}= \frac{1}{2\pi }\int dr \left[ \frac{\omega }{
N^{2}} \left( \frac{X_{1}}{Z_{1}}-\frac{X_{2}}{Z_{2}} \right) + \frac{1}{N
\sqrt{g_{\phi }}} \left( \frac{X_{1}}{Z_{1}}+\frac{X_{2}}{Z_{2}}\right)
\right] .  \label{max}
\end{equation}
For a nonextremal black hole it is finite and gives the unconditional upper
bound.

\section{Extremal black hole\label{extremal}}

By definition, this means
\begin{equation}
N^{2}\approx D(r-r_{+})^{2} ,
\label{44dop}
\end{equation}
the constant $D>0$.

A usual particle reaches the horizon for a finite proper time $\tau $.
In doing so, we see that in~(\ref{fr}) the first term dominates,
\begin{equation}
n\approx \frac{\omega_{H}}{2\pi D}(r_{1}-r_{+})^{-1}+\frac{B_{1}}{D}
\left\vert \ln (r_{1}-r_{+})\right\vert +O(1)\rightarrow \infty .
\label{extr}
\end{equation}
For two different particles, it follows from~(\ref{12}), (\ref{xx}) that
$ n_{1}-n_{2}$ remains finite similarly to the nonextremal case.

For the critical particle, the situation is more special. Typically, as is
well known, if a particle approaches a black hole, the time $t$ measured by
a remote observer at infinity diverges while the proper time remains finite.
By contrast, if the critical particle approaches the extremal horizon, not
only $t$ but also the proper time $\tau $ diverges. It is this circumstance
that requires considering this case separately and needs some care.

It follows from~(\ref{Z}), (\ref{exp})  (\ref{44dop}) that
\begin{equation}
X\approx B_{1}L(r-r_{+}),
\end{equation}
\begin{equation}
Z\approx \sqrt{\left( B_{1}^{2}-\frac{D}{g_{\phi }}\right) L^{2}-m^{2}D}
\,(r-r_{+}).  \label{zb}
\end{equation}
According to~(\ref{lp}), both terms in~(\ref{fr}) are positive and
\begin{equation}
n\approx \frac{\omega_{H}B_{1}L}{2\pi D\sqrt{\left( B_{1}^{2}-\frac{D}{
g_{\phi }}\right) L^{2}-m^{2}D}}(r_{1}-r_{+})^{-1}\rightarrow \infty .
\label{ncr}
\end{equation}
Thus in all cases when a particle reaches the horizon, $\phi \rightarrow
+\infty $, $n\rightarrow \infty $. Also, $\tau \rightarrow \infty $.

One can also consider two different critical particles. Then, it is easy to
obtain that, since $n$ depends on $L$,
$n_{1}-n_{2}=O \left((r_{1}-r_{+})^{-1}\right)$ \, still diverges.
Eq.~(\ref{max}) now gives no upper bound since $n_{\max }$ diverges due to
the second term.

It is worth to pay attention to the following difference between properties
of $n$ in the flat spacetime and black hole background.
We saw in Sec.~\ref{flat} that in the first case, $n_{1}-n_{2}$ tends
to infinity as a function of $r$ when $r\rightarrow \infty $.
However, if we consider $n_{1}-n_{2}$ as
a function of $t$, $n_{1}-n_{2}$ stays finite even if $t\rightarrow \infty $.
This crucial difference arises due to the fact that, according to~(\ref{b})
and~(\ref{rb}), for large $t$ and $r$ the radial coordinate depends on
particle's parameters. Therefore, the situations when either $r$ or $t$ are
equal for both particles are inequivalent. By contrast, in the black hole
case, $r\rightarrow r_{+}$ for any particle when $t\rightarrow \infty $, so
the limit is universal and there is no sense to distinguish the situations
for a given $r$ and given~$t$. This is valid both for nonextremal and
extremal black holes.

\section{What does a falling observer see?\label{fall}}

Any measurements imply the presence of some frame realized by an observer
with corresponding devices. However, before reaching the horizon, a falling
observer crosses the boundary of a region where
\begin{equation}
g_{00}=0 .
\end{equation}
Inside it, $g_{00}>0$. This is called the ergoregion or ergosphere (see,
e.g. Sec.~3.3.2 of~\cite{fn} or Ch.~33.2 -- 33.4 of~\cite{grav}). As a
consequence, motion with $\phi =\mathrm{const}$ is impossible inside the
ergoregion since otherwise a wordline would become spacelike, so any
particle rotates. Therefore, any frame rotates with respect to remote stars
as well. The physical angular velocity should be calculated by subtracting
the velocity of space itself from it. There are two general methods to
realize this procedure.

\subsection{Reference particle}

The first one consists in that in any point of the trajectory of particle 1
we use the reference particle 0 with given energy $E_{0}$ and angular
momentum $L_{0}$ and find the rotation angle with respect to the observer
comoving with particle 0. Afterwards, we integrate along the path of
particle 1.

In other words, we use~(\ref{om}) for both particles with the same $dt$ and
use eq.~(\ref{rt}) for particle 1 to integrate over $dr$.
Then, instead of~(\ref{12}), we have
\begin{eqnarray}
n_{1}-n_{0}=\frac{\Delta \phi }{2\pi } &= & \frac{1}{2\pi }\int \frac{dtN^{2}
}{ g_{\phi }} \left( \frac{L_{1}}{X_{1}}-\frac{L_{0}}{X_{0}} \right)= \frac{1
}{2\pi } \int_{r_{1}}^{r_{0}}\frac{drX_{1}}{g_{\phi }Z_{1}} \left( \frac{
L_{1}}{X_{1}}-\frac{L_{0}}{X_{0}} \right) =  \nonumber \\
&& \frac{1}{2\pi }\int_{r_{1}}^{r_{0}}\frac{dr}{g_{\phi }}(\frac{
L_{1}E_{0}-L_{0}E_{1}}{Z_{1}X_{0}}) .  \label{10}
\end{eqnarray}
Obviously, $n_{1}-n_{0}$ remains finite for a usual particle. For the
critical particle moving near the extremal black hole, it follows
from~(\ref{zb}) that $n_{1}-n_{0}$ diverges logarithmically.

Using~(\ref{maxmin}), it is easy to obtain the upper bound similar
to~(\ref{max}):
\begin{equation}
n_{1}-n_{0}<n_{\max }^{(10)}=\frac{1}{\pi }\int \frac{drX_{1}}{\sqrt{g_{\phi
}}Z_{1}N} .  \label{max10}
\end{equation}

\subsection{Rotating frame}

The second method consists in that we choose some rotating frame with a
prescribed angular velocity $\Omega (r),$ so that
\begin{equation}
\left( \frac{d\phi }{dt}\right)_{0}=\Omega .
\end{equation}
Then, by construction, we define the quantity
\begin{equation}
n_{1}-n_{\Omega }=\frac{1}{2\pi }\int_{r_{1}}^{r_{0}} \left( \frac{(\omega -
\Omega )X}{ZN^{2}}+\frac{L}{g_{\phi }Z} \right) dr .
\label{omom}
\end{equation}
There are different possible choices of such a frame considered below.

\section{System corotating with a black hole\label{corot}}

Now,
\begin{equation}
\Omega =\omega_{H}
\end{equation}
coincides with the angular velocity of a black hole.
Then, we have from~(\ref{omom}) the number of revolution with respect to
a black hole. Instead if~(\ref{fr}), we have
\begin{equation}
n_{1}-n_{\mathrm{bh}}=\frac{1}{2\pi }\int_{r}^{r_{0}}dr \left[ \frac{(\omega
- \omega_{H})X}{ZN^{2}}+\frac{L}{g_{\phi }Z} \right] ,  \label{psin}
\end{equation}%
where we used subscript ``$\mathrm{bh}$'' to stress a role of a black hole.

Let $r\rightarrow r_{+}$. We will discuss different cases separately.

\subsection{Nonextremal black hole}

Both terms in~(\ref{psin}) are finite, so $n_{1}-n_{\mathrm{bh}}$ is finite
as well. A particle makes a finite number of revolution before falling into
a black hole.

\subsection{Extremal black hole}

For a usual particle, the second term in~(\ref{psin}) is finite. However,
taking into account~(\ref{omega}), we see that the first term gives us
\begin{equation}
n_{1}-n_{\mathrm{bh}}=O(\ln (r-r_{+}))\rightarrow -\infty .
\end{equation}
For the critical particle, taking into account~(\ref{exp}) with $X_{H}=0$
and~(\ref{zb}), we have from~(\ref{psin})
\begin{equation}
n_{1}-n_{\mathrm{bh}}\approx \frac{\left( B_{1}^{2}-\frac{D}{g_{\phi }}
\right) L}{2\pi D\sqrt{ \left( B_{1}^{2}-\frac{D}{g_{\phi }} \right)
L^{2}-m^{2}D}} \ln (r-r_{+}) \rightarrow -\infty
\end{equation}
since $B_{1}^{2}-\frac{D}{g_{\phi }}>0$. Otherwise, a particle cannot
approach the horizon.

\section{Rotation with respect to local ZAMO observer\label{nrf}}

Now, we choose
\begin{equation}
\Omega =\omega .
\end{equation}
    This corresponds to a so-called zero-angular momentum observers~\cite{72}
(ZAMO) as a reference particle 0 in any point.
    It follows from~(\ref{omom}) that
\begin{equation}
n-n_{\mathrm{zamo}}=\frac{L}{2\pi }\int_{r_{1}}^{r_{0}}\frac{dr}{Zg_{\phi }}=
\frac{L}{2\pi m}\int \frac{d\tau }{g_{\phi }} .
\label{zamo}
\end{equation}
This formula can be obtained also from~(\ref{10}), if one puts $L_{0}=0$.

    As, by assumption, $g_{\phi }$ is finite, we see that in the limit
$ r_{1}\rightarrow r_{+}$ the quantity $n-n_{\mathrm{zamo}}$ is finite
for a usual particle as well as the proper time $\tau $ (\ref{pr}).
    This holds both for nonextremal and extremal black holes.
        For the critical one moving in the background of the extremal black hole,
they both are infinite.

\section{Intermediate case}

Now, we can define
\begin{equation}
\Omega =\omega +\frac{N}{\sqrt{g_{\phi }}}\alpha ,  \label{al}
\end{equation}
where $-1\leq \alpha \leq 1$. In the particular case $\alpha =0$, we return
to the ZAMO frame. If $\alpha =\pm 1$, $\Omega =\Omega_{\pm }$ defined
in~(\ref{pm}). Now, we obtain from~(\ref{omom})
\begin{equation}
n_{1}-n_{\alpha }=\frac{1}{2\pi }\int_{r_{1}}^{r_{0}} \left( -\frac{\alpha X
}{\sqrt{g_\phi}\, N}+ \frac{L}{g_{\phi }} \right) \frac{dr}{Z} .
\end{equation}

\subsection{Nonextremal black hole}

Now $n_{1}-n_{\alpha }$ is finite.

\subsection{Extremal black hole}

For a usual particle,
\begin{equation}
n_{1}-n_{\alpha }\approx \frac{\alpha \ln (r-r_{+})}{2\pi \sqrt{D}} .
\end{equation}%
It tends to $-\infty $ or $+\infty $ depending on $\alpha $.

For the critical particle,%
\begin{equation}
n_{1}-n_{\alpha }\approx -\frac{L}{2\pi g_{\phi }\sqrt{ \left( B_{1}^{2} -
\frac{D}{g_{\phi }} \right) L^{2}-m^{2}D}} \ln (r-r_{+}) \left(
1-B_{1}\alpha \frac{\sqrt{g_{\phi }}}{\sqrt{D}} \right) .  \label{acrit}
\end{equation}
It diverges and, depending on $\alpha $, can tend to either $+\infty $ or
$ -\infty $.

\section{Universality of dragging effect and nature of a black hole}

The above results can be summarized in the following Table~1.
\newline
\newline
\begin{tabular}{|p{0.26\textwidth}|p{0.099\textwidth}|p{0.099\textwidth}|p{0.099
\textwidth}|p{0.11\textwidth}|p{0.099\textwidth}|}
\hline
& $n_{1}$ & $n_{1}\!-\!n_{2}$ & $n_{1}\!-\!n_{\mathrm{bh}}\!$ & $\! n_{1}
\!-\! n_{\mathrm{zamo}}\!$ & $n_{1} \!-\! n_{\alpha }$ \\ \hline
Flat & Infinite & Infinite & - & - & - \\ \hline
Nonextremal black hole & Infinite & Finite & Finite & Finite & Finite \\
\hline
Extremal black hole (usual particles) & Infinite & Finite & Infinite & Finite
& Infinite \\ \hline
Extremal black hole (critical particles) & Infinite & Infinite & Infinite &
Infinite & Infinite \\ \hline
\end{tabular}
\vspace{4pt}
\newline

{\bf Table~1}. Properties of numbers of revolutions for different types of
particles in different backgrounds.\newline
\newline
Here, $n_{1}$ refers to a number of revolutions for one particle measured by
a remote observer, $n_{1}-n_{2}$ represents the corresponding difference for
two generic different particles. The quantity $n_{1}$ diverges always (in a
similar way, say, the gravitational redshift diverges when measured by a
remote observer independently of a type of a black hole).
The quantity $ n_{1}-n_{\mathrm{bh}}$ gives a number or revolutions of
particle~1 defined according to subsection~\ref{corot} with respect to
a black hole.
The quantity $n_{1}-n_{\mathrm{zamo}}$ gives the corresponding quantity with
respect to ZAMO observers according to the definitions in
subsection~\ref{nrf}.

We saw that for a usual particle, according to~(\ref{nd}), (\ref{extr})
the divergent terms in $n$ have an universal form in the sense that
they do not contain the energy or angular momentum of a particle itself
and depends on characteristics of a black hole only.
    This is valid both for nonextremal and
extremal black holes. In the first case they are milder and contain only the
logarithmic term. As divergences are universal, a subtraction procedure
connected with the comparison of rotation of two particles, kills
divergences, the rest remaining finite. However, if, instead, we measure
particle rotation with respect to rotation of a black hole, the answer
depends on the rate with which the coefficient $\omega $ responsible for
dragging effects approaches $\omega_{H}$. For nonextremal black holes
$ \omega -\omega_{H}=O(N^{2})$ \cite{dirty} and in the first term
of~(\ref{psin}) the numerator has the same order as the denominator,
so the outcome is finite.
For nonextremal black holes $\omega -\omega_{H}=O(N)$ and goes
to zero more slowly than the denominator that gives an infinite result
in $ n_{1}-n_{bh}$ according to the table.
The same reasoning apply to $ n_{1}-n_{\alpha }$.
Thus the divergences under discussion arise not only due
to rotation of space itself because of a strong dragging effect in the
ergoregion, but also due to the fact that this effect is especially strong
for extremal black holes.

For the critical particle near the extremal black hole, the divergences are
not universal and contain characteristics of a particle as is seen
from~(\ref{ncr}).
Therefore, cancellation is, in general, impossible and the result
depends on a procedure and divergences persist, as this is seen from the
last line of Table~1.

\section{Kerr metric}

Now, we illustrate the general properties discussed above using the Kerr
metric as an example. It is very important for astrophysics, represents an
exact solution and admits detailed analysis of particle motion, even for the
nonequatorial case.

\subsection{Form of metric and equations for geodesics}

In the Boyer-Lindquist coordinates it reads (see, e.g. Sec.~61, 62 in Ch.~7
of~\cite{ch} or Sec.~3.4.1 of~\cite{fn})
\begin{equation}
ds^{2}=-\frac{\rho^{2}\Delta }{\Sigma^{2}}dt^{2}+ \frac{\sin^{2}\theta }{%
\rho^{2}}\Sigma^{2}(d\phi -\omega dt)^{2}+ \frac{\rho^{2}}{\Delta }
dr^{2}+\rho ^{2}d\theta^{2} ,
\end{equation}
where
\begin{equation}
\rho^{2}=r^{2}+a^{2}\cos^{2}\theta , \ \ \ \Delta =r^{2}-2Mr+a^{2} ,
\end{equation}
\begin{equation}
\Sigma^{2}=(r^{2}+a^{2})^{2}-a^{2} \Delta \sin^{2}\theta , \ \ \ \omega =
\frac{2Mra}{\Sigma^{2}} ,
\end{equation}%
$M$ is a black hole mass, $aM$ being its angular momentum.
We assume that $ 0\leq a\leq M$.
The event horizon lies at $r= r_{+} \equiv M + \sqrt{M^{2}-a^{2}} $.
On the boundary of the ergoregion
$r= r_{1} \equiv M + \sqrt{M^{2}-a^{2} \cos^2 \theta} $
that defines the limit of stationarity lies the quantity
\begin{equation}
S(r,\theta ) = r^{2}-2Mr + a^{2}\cos^{2}\theta
\end{equation}
vanishes. Inside the ergosphere, $S<0$.

The equations for geodesics in the Kerr metric read
\begin{equation}
\rho^{2}\frac{dt}{d\lambda }=\frac{1}{\Delta }(\Sigma^{2}E-2MraL) , \ \ \ \
\rho^{2}\frac{d\phi }{d\lambda }=\frac{1}{\Delta } \left( 2MraE+\frac{SL}{
\sin^{2}\theta } \right) ,
\label{tphlam}
\end{equation}
\begin{equation}
\rho^{2}\frac{dr}{d\lambda }=\sigma_{r}\sqrt{R} , \ \ \ \ \
\rho^{2}\frac{d\theta}{d\lambda }=\sigma_{\theta}\sqrt{ \Theta} ,
\end{equation}
\begin{equation}
R=\Sigma^{2}E^{2}-\frac{SL^{2}}{\sin^{2}\theta }-4MraLE-\Delta (m^{2}
\rho^{2}+\Theta ) ,
\end{equation}
\begin{equation}
\Theta =Q-\cos^{2}\theta \left[ a^{2}(m^{2}-E^{2})+
\frac{L^{2}}{\sin^{2}\theta } \right] .
\end{equation}
Here we use notations $E$ and $L$ from Sec. II, $m$ is a particle's mass,
$ \lambda $ being the affine parameter along the geodesics. For a massive
particle, $\lambda =\tau /m$. $Q$~is the Carter constant. For the equatorial
motion ($\theta =\pi /2$), $Q=0$. The factors $\sigma_{r}=\pm 1$ and
$ \sigma_{\theta }=\pm 1$ define the direction of motion along the
coordinates~$r$ and~$\theta $.
The angular velocity can be found from~(\ref{tphlam})
\begin{equation}
\frac{d\phi }{d t }= \omega +\frac{\Delta \rho^{4}L}{ \Sigma^{2}(
\Sigma^{2}E-2MraL) \sin^{2}\theta} ,
\end{equation}
The quantities~(\ref{pm}) are equal to
\begin{equation}
\Omega_{\pm } = \omega \pm \frac{\rho^{2}\sqrt{\Delta }}{\Sigma^{2} \sin
\theta } .  \label{omk}
\end{equation}

\begin{figure}[th]
\centering
\includegraphics[width=6cm]{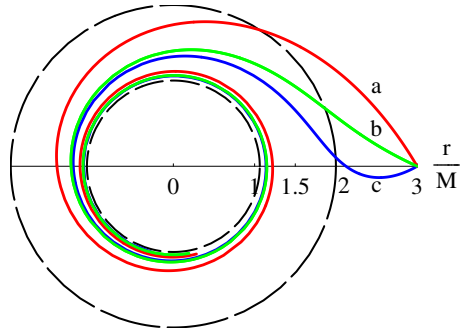}
\caption{The trajectories (a) of the particle with mass $m$ and $L=1.9 mM$,
(b) a photon with $L=0$, (c) the particle with $L=-4 mM$.}
\label{Fig0}
\end{figure}

On Fig.~\ref{Fig0} we shown the trajectories of particles free falling in
equatorial plane into black hole with $a = 0.998 M$ from point $r= 3M$.
Into the ergosphere (within of dashed line with $r=2M$) rotation of all
particles, irrespective of sign of a projection of the angular momentum, is
directed towards rotation of the black hole. The trajectories of particles
wrap an infinite number of times around the event horizon when they approach
it.

\subsection{Relative number of revolutions}

For simplicity, we restrict ourselves to the equatorial motion and consider
the quantity $n_{1}-n_{0}$. Then, after substitution of the explicit
expressions for the Kerr metric into eq.~(\ref{10}), we have

\begin{equation}
n_{1}-n_{0}=\frac{1}{2\pi }\int dr\rho^{4}\frac{(E_{0} L - E L_{0})}{
\sqrt{R}\,(\Sigma^{2}E_{0}-2MraL_{0}) } .
\label{kerr}
\end{equation}
In the horizon limit $r\rightarrow r_{+}$, $\Sigma^{2}\rightarrow
4M^{2}r_{+}^{2}$, $R\rightarrow (2Mr_{+}E-aL)^{2}$. Therefore, for all
particles falling into a nonextremal black hole, the rotation angle
$\Delta \phi $ remains finite.
It is also finite in the case of the extremal black
hole, except for the critical particle for which $\Delta \phi $ diverges as
well as the corresponding proper time in accordance with general statements
made above.

For given values of parameters, the integral~(\ref{kerr}) can be calculated
numerically. Let, for example, $E=m$, so a particle starts its motion from
infinity with the zero velocity. Then, it is known~\cite{fn}, Sec.~3.4.5,
that it can reach the horizon only for the values of the angular momentum in
the range
\begin{equation}
-2 \left( 1+\sqrt{1+\frac{a}{M}}\, \right)\leq \frac{L}{mM} \leq 2
\left( 1+ \sqrt{1-\frac{a}{M}}\, \right) .
\end{equation}
Let, say, $L=mM$ and $L_{0}=0$.
Then, it turns out that 
the number of revolutions $n_{1}-n_{0}\sim 1$.

It follows from~(\ref{max10}) that
\begin{equation}
\left( \Delta \phi \right)_{\max }=\int dr \frac{2\rho^{2}(E
\Sigma^{2}-2MraL)}{\Sigma^{2} \sqrt{\Delta R}} .
\end{equation}
This integral converges in the nonextremal case and diverges in the extremal
one.

\begin{figure}[th]
\centering
\includegraphics[width=6cm]{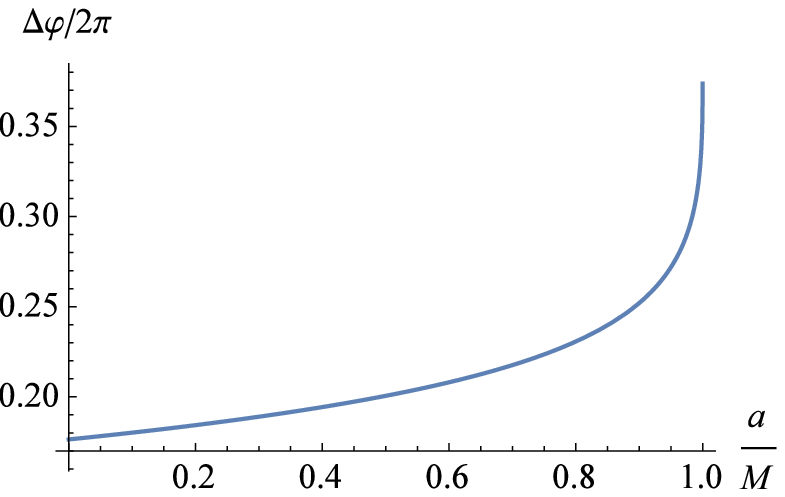}
\caption{Relative number of revolutions for two falling particles.}
\label{Fig1}
\end{figure}

On Fig.~\ref{Fig1} we depict $n_{1}-n_{0}$ for the case when a particle has
$ E=m$ and $L=1.9mM$, $L_{0}=0$.
We take $r_{0}=9M$. We see that angle of rotation is relatively small, it is
less than 1/3 of the full revolution.

\begin{figure}[th]
\centering
\includegraphics[width=6cm]{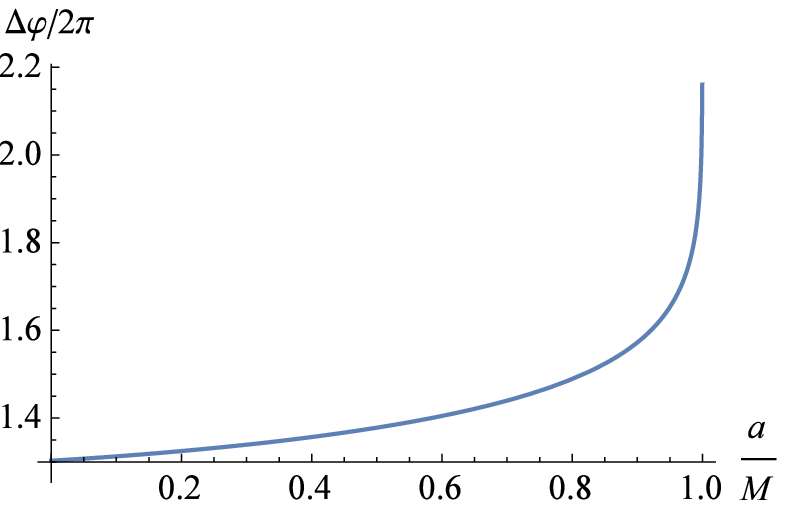}
\caption{Maximal relative number of revolutions.}
\label{Fig2}
\end{figure}

On Fig.~\ref{Fig2} we depict the upper bound on the number of revolution
$ n_{\max }=\left( \Delta \phi \right)_{\max } / (2\pi )$ for the same
particle as on Fig.~\ref{Fig1}.
The quantity $\left( \Delta \phi \right)_{\max }$ is again calculated
starting form $r=9M$.
For the Thorne limit~\cite{th} $a/M=0.998$ the number of revolutions $n
\approx 2$. For the extremal black hole ($a=M$) $n\rightarrow \infty $
in accordance with the general results of Sections~(\ref{extremal})
and~(\ref{fall}).

\section{Conclusion}

When one speaks about the number of revolutions $n$ of a particle around a
black hole, one should state clearly which kind of observers is used. In
doing so, it is necessary to distinguish between two physically inequivalent
cases since the result and the very notion of $n$ depends on a reference
object with respect to which we counter revolutions. For a remote observer a
natural coordinate is a Boyer-Lindquist one that approaches a usual angle
variable typical of the Minkowski metric. Then, such an observer defines
revolutions with respect to the line $\phi =\mathrm{const}$ and finds an
infinite~$n$. The main reason of such divergences is a strong dragging
effect because of which space itself rotates inside the ergoregion. This
phenomenon reveals itself in somewhat different way for nonextremal and
extremal black holes.

In the first case, it is (i) universal in that the divergent term does not
depend on characteristics of particles and (ii) contains only logarithmic
divergences. Another observer who rotates near a black hole sees that a
reference objects rotates as well. If this observer calculates $n$, he
subtracts the contribution of the frame from his own rotation. In other
words, the quantity $n$ is regularized. It follows from the aforementioned
properties~(i) and (ii) that divergences inherent to an observer and the
frame mutually cancel and the outcome is finite.

In the second (extremal) case, the situation depends also on a type of a
particle and the kind of regularization, so an infinite number of
revolutions measured by a local observer becomes possible.

The results of the present work can, in principle, be useful for the
analysis of the behavior of clouds of particles in the course of
gravitational collapse on rotating black holes.

\begin{acknowledgements}
    This work was supported by the Russian Government Program of
Competitive Growth of Kazan Federal University.
    The work of Yu.\,P. was supported also by the Russian Foundation for
Basic Research, grant No. 15-02-06818-a.
    The work of O.\,Z. was also supported by SFFR, Ukraine, Project No. 32367.
\end{acknowledgements}


\end{document}